\pdfoutput=1
\documentclass{llncs}
\usepackage{graphicx}

\begin{document}
\newcommand{\vpi}{\mathbf{\pi}}
\newcommand{\rar}{\rightarrow}
\newcommand{\lar}{\leftarrow}
\newcommand{\ra}{\rangle}
\newcommand{\la}{\langle}
\newcommand{\ttt}{\texttt}
\newcommand{\mbb}{\mathbb}
\newcommand{\mbf}{\mathbf}
\newcommand{\mca}{\mathcal}

\frontmatter
\pagestyle{headings} 
\addtocmark{Computational Eudaemonics}

\mainmatter

\title{Faith in the Algorithm, Part 2: \\ Computational Eudaemonics}

\titlerunning{Computational Eudaemonics}  

\author{Marko A. Rodriguez\inst{1} \and Jennifer H. Watkins\inst{2}}

\authorrunning{Marko A. Rodriguez and Jennifer H. Watkins}
\tocauthor{Marko A. Rodriguez and Jennifer H. Watkins}

\institute{Center for Nonlinear Studies \\ Los Alamos National Laboratory, Los Alamos NM 87545, USA \\
\email{marko@lanl.gov}
\and
International and Applied Technology \\ Los Alamos National Laboratory, Los Alamos NM 87545, USA \\
\email{jhw@lanl.gov}
}

\maketitle

\begin{abstract}
Eudaemonics is the study of the nature, causes, and conditions of human well-being. According to the ethical theory of eudaemonia, reaping satisfaction and fulfillment from life is not only a desirable end, but a moral responsibility. However, in modern society, many individuals struggle to meet this responsibility. Computational mechanisms could better enable individuals to achieve eudaemonia by yielding practical real-world systems that embody algorithms that promote human flourishing. This article presents eudaemonic systems as the evolutionary goal of the present day recommender system.
\end{abstract}

\begin{quote}
\textit{[Those who condemn individualism] slur over the chief problems---that of remaking society to serve the growth of a new type of individual.}
 \begin{flushright}
 	John Dewey, ``Individualism Old and New"
\end{flushright}
\end{quote}

\section{Introduction}

Eudaemonia is the theory that the highest ethical goal is personal happiness and well-being \cite{nico:aristotle350}. This theory holds that an ethical life is one filled with the meaning and satisfaction that arises from living according to one's values---where everything one does is of great importance to their character. Eudaemonia parallels the notion of Abraham Maslow's self-actualization \cite{motivation:maslow} and Mih\'{a}ly Cs\'{i}kszentmih\'{a}lyi's flow state \cite{csi:flow1990} except that, as an ethical theory, it argues that it is a personal responsibility to strive for this state. As a social theory, eudaemonia holds that the purpose of society is to promote this state in all of its people. The ethical foundation of personal flourishing is grounded in the contention that the purpose of life is to reap satisfaction and fulfillment from an engagement in the world and that such a state is objectively good for society. Thus, learning how to flourish is a form of moral development.

Moral development, when used in this sense, extends beyond civility, honesty, and other facets of rectitude. It refers to a personal onus to achieve well-being. One proponent of the ethical theory of eudaemonia, David L. Norton, states that ``[...] the broader eudaimonistic thesis is that all virtues subsist \textit{in potentia} in every person; thus to be a human being is to be capable of manifesting virtues, and the problem of moral development is the problem of discovering the conditions of their manifestation" \cite{demo:norton1995}. Typically, the discovery of the conditions that will manifest virtues in the individual is guided by the recommendations of family, friends, and community---those who know the individual well and the options available to them. Despite this guidance, the achievement of eudaemonia remains elusive for most. Maslow notes that a very small group of people achieve self-actualization and Cs\'{i}kszentmih\'{a}lyi has shown that very few are able to control their consciousness well enough to reliably reach the state of flow. Given the individual moral imperative to achieve eudaemonia and the resulting societal benefits, resources should be dedicated to guaranteeing this realization for as many people as possible.

Eudaemonics is the study of the nature, causes, and conditions of eudaemonia \cite{hard:flanagan2007}. For Owen Flanagan, the domains of moral and political philosophy, neuroethics, neuroeconomics, and positive psychology are the sources from which a developed understanding of human well-being will spring. In this article, it is posited that \textit{computational eudaemonics} will make advances to bring eudaemonia to more than a select few in society. Computer and information science can greatly contribute to the eudaemonic endeavor by yielding practical real-world systems that embody algorithms that promote human flourishing. Systems that promote eudaemonia are called eudaemonic systems. Such systems would foster eudaemonia by providing the right conditions for the manifestation of virtues. This article presents a vision of eudaemonic systems as the evolutionary goal of the present day recommender system.

\section{From Recommender to Eudaemonic Systems}

The purpose of a eudaemonic system is to produce societies in which the individuals experience satisfaction through a deep engagement in the world. This engagement can be fostered by uniting individuals with those resources that resonate with their nature. Resources can take many forms, a few of which are itemized below.
\begin{itemize}
	\item activities: vocations, hobbies, gatherings, projects.
	\item education: universities, lectures, areas of study.
	\item entertainment: books, movies, music, shows.
	\item people: friends, work associates, life partners.
	\item places: to live, to vacation, to dine.
\end{itemize}

There are many ways a eudaemonic system could contribute to individual well-being. Perhaps the most ambitious eudaemonic system is one that supplies the satisfaction of the need for a resource before the need is even felt. For Thomas Hobbes, eudaemonia is encumbered by conation---goals, plans, and desires \cite{hobbes:leviathan1651}. Practically speaking, humans seek books and movies to stimulate their cognitive faculties, friends and partners to fulfill their social affinities, art to entice their affective natures, and sports to satiate their physical needs. While every individual longs for varying degrees of these requirements, in general, a flourishing life is one where all these requirements are met through the active process of enacting them \cite{good:kraut2007}. Thus, a Hobbesian eudaemonic system would be one that satisfied requirements before they were felt (pre-conation), so that the experience of need could not disrupt a life of contentment. Through computational mechanisms, it may be possible to produce pre-conate eudaemonic systems. A pre-conate system is one that makes use of indicators of coming discontent and provides avenues to rectify the situation prior to its actualization. 

Recommender systems \cite{recommned:resnick1997}, when viewed within the context of the eudaemonic thesis, could evolve to become such systems. A recommender system is an information filtering tool that matches individuals to resources of potential interest. Such systems are commonly employed by businesses in an attempt to sell more products. However, this conceptualization of the recommender system trivializes their potential role. 

The satisfaction one reaps from the world can be represented in terms of one's interactions with resources. These interactions need not be extraordinary, but are the stuff of everyday life. Norton articulates the importance of everyday activities when he states that ``if the development of character is the moral objective, it is obvious that [...] the choices of vocation and avocations to pursue, of friends to cultivate, of books to read are moral for they clearly influence such development" \cite{demo:norton1995}. For the techno-social society, this development of character is driven every day, to some extent, by the use of recommender systems. Thus, to the extent that recommender systems influence choices, they already influence moral development. By purposely designing these systems to orient individuals toward life optima, recommender systems can evolve to become eudaemonic systems.

\begin{figure*}[h!]
	\centering
		\includegraphics[width=1.0\textwidth]{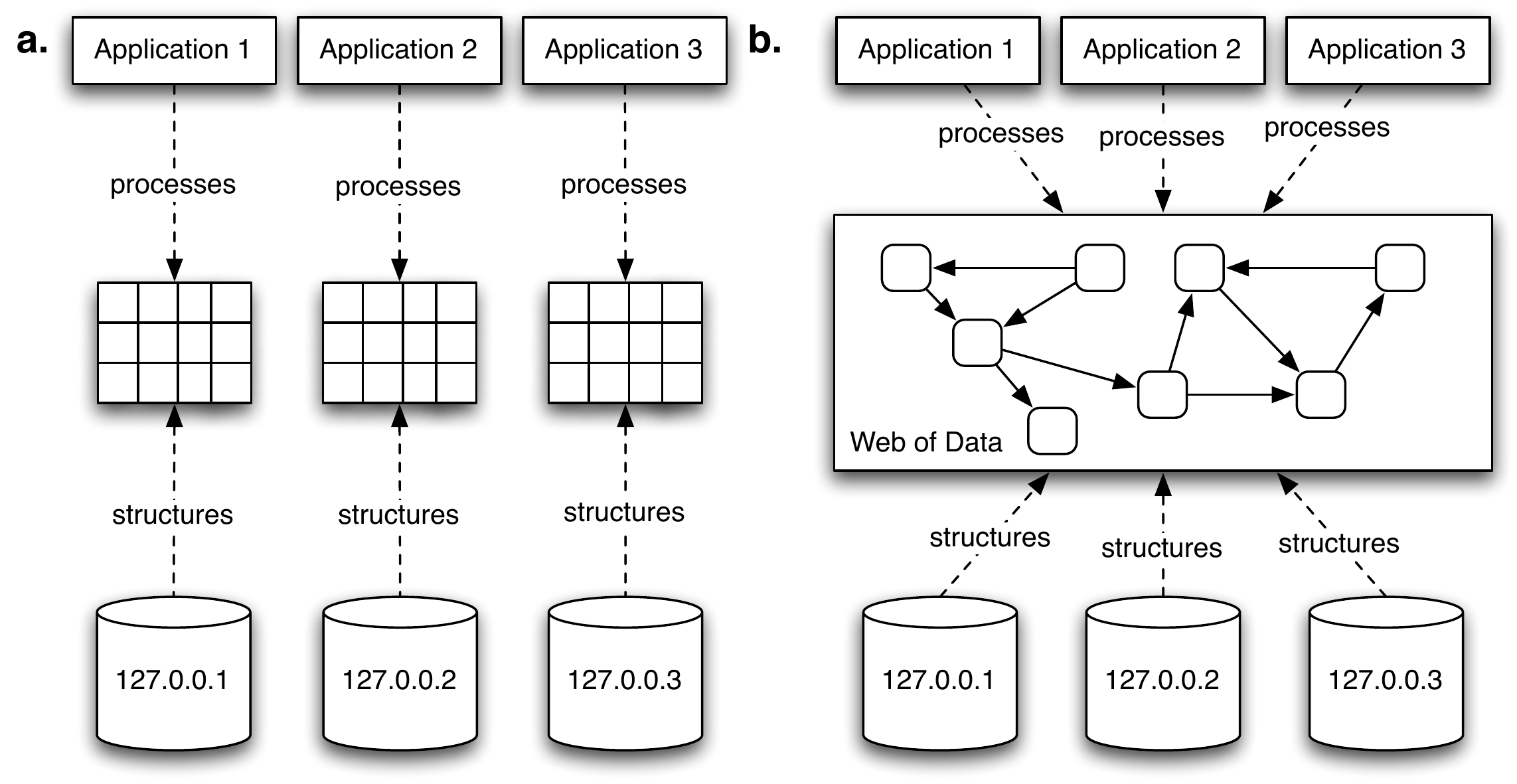}
	\caption{a.) The current paradigm in which the application and the data upon which it operates are tightly coupled both technically and proprietarily---\S \ref{sec:rec}. b.) The emerging Web of Data provides a collectively generated, publicly accessible world model that can be leveraged by independent application developers---\S \ref{sec:eud}.\label{fig:structures-processes}}
\end{figure*}

The current generation of recommender systems are limited to a particular representational slice of the world (such as movies). This is represented in Figure \ref{fig:structures-processes}a, where there exists a tight coupling between the data and the application which operates on that data. A eudaemonic system must account not for a single aspect of an individual's life, but for the multitude of domains in which that individual exists. The emerging Web of Data provides a distributed data structure that cleanly separates the data providers from the application developers. This is represented in Figure \ref{fig:structures-processes}b. The remainder of this section will discuss recommender systems and their transition to eudaemonic systems through the exploitation of the Web of Data.

\subsection{Recommender Systems}\label{sec:rec}

Most recommender systems model individual users, resources, and their relationships to one another \cite{recommned:resnick1997}. For example, in an online store, users may have an \ttt{ex:hasPurchased} relationship to some of the store's products. If the purchasing behavior of user $x$ and user $y$ has a strong, positive correlation, then any products purchased by only one can be recommended to the other. Purchasing behavior is not the only way in which resources are deemed similar. It is possible to relate resources by shared metadata properties \cite{content:pazzani2007}. For example, an online movie rental service can represent movie $a$ as having an \ttt{ex:directedBy} relationship to director $b$ and director $b$ can maintain an \ttt{ex:directed} relationship to movie $c$. The similarity that exists between movies $a$ and $c$ is determined, not by user behavior, but by similarity of metadata---the same person directed both. By building a graph of typed relationships between resources, it is possible to identify different forms of relatedness and utilize these forms to aid an individual in their decision making process regarding the use of such resources.

The power of recommender systems is currently limited because they rely on a single silo of data that must be generated before they can provide useful recommendations (see Figure \ref{fig:structures-processes}a). Due to the data acquisition hurdle, application designers must focus on a particular niche in which to provide recommendations. For example, services either provide recommendations for books,\footnote{For example: Amazon.com, Feedbooks.com} or for music,\footnote{For example: Pandora.com, Last.fm}, or for partners,\footnote{For example: Match.com, Chemistry.com, eHarmony.com} etc. With such a limited worldview, these services do not respect the multi-faceted nature of human beings. If a system only has access to data on movies, then it can never recommend the perfect beach novel. Eudaemonia requires a complete representation of the domains in which one conducts life in order to recommend the right resource at the right time. Therefore, eudaemonic systems require an integrated representation of the world's resources and the individual's place within them.

\subsection{Eudaemonic Systems}\label{sec:eud}

The recommender system data structure described previously can be conveniently represented as a multi-relational network. The most prevalent multi-relational data model is the Resource Description Framework (RDF) of the Semantic Web initiative. The Semantic Web's Linked Data community is dedicated to the development of the emerging RDF-based Web of Data. On the Web of Data, all data is represented in the URI address space and interlinked to form a single, global data structure that can be used by both man and machine for various application scenarios (see Figure \ref{fig:structures-processes}b) \cite{linkeddata:bizer2008}.\footnote{The public exposure of data has stimulated interest in the development of the legal structures for the use of such data. Much like the Open Source movement, the Linked Data community is actively involved in the Open Data movement \cite{opendata:miller2008}.} The Web of Data provides two significant benefits over the data silos used by recommender systems. First, application developers need not focus on data acquisition and instead can focus directly on algorithm development. This feature ultimately reduces the labor involved in web service deployment. Second, the application developer can create algorithms that make use of a rich world model that incorporates the various ways in which resources relate to each other. Thus, these algorithms have a larger knowledge-base with which to understand the world and the individual's place within it.

\begin{figure*}[h!]
	\centering
		\includegraphics[width=1.0\textwidth]{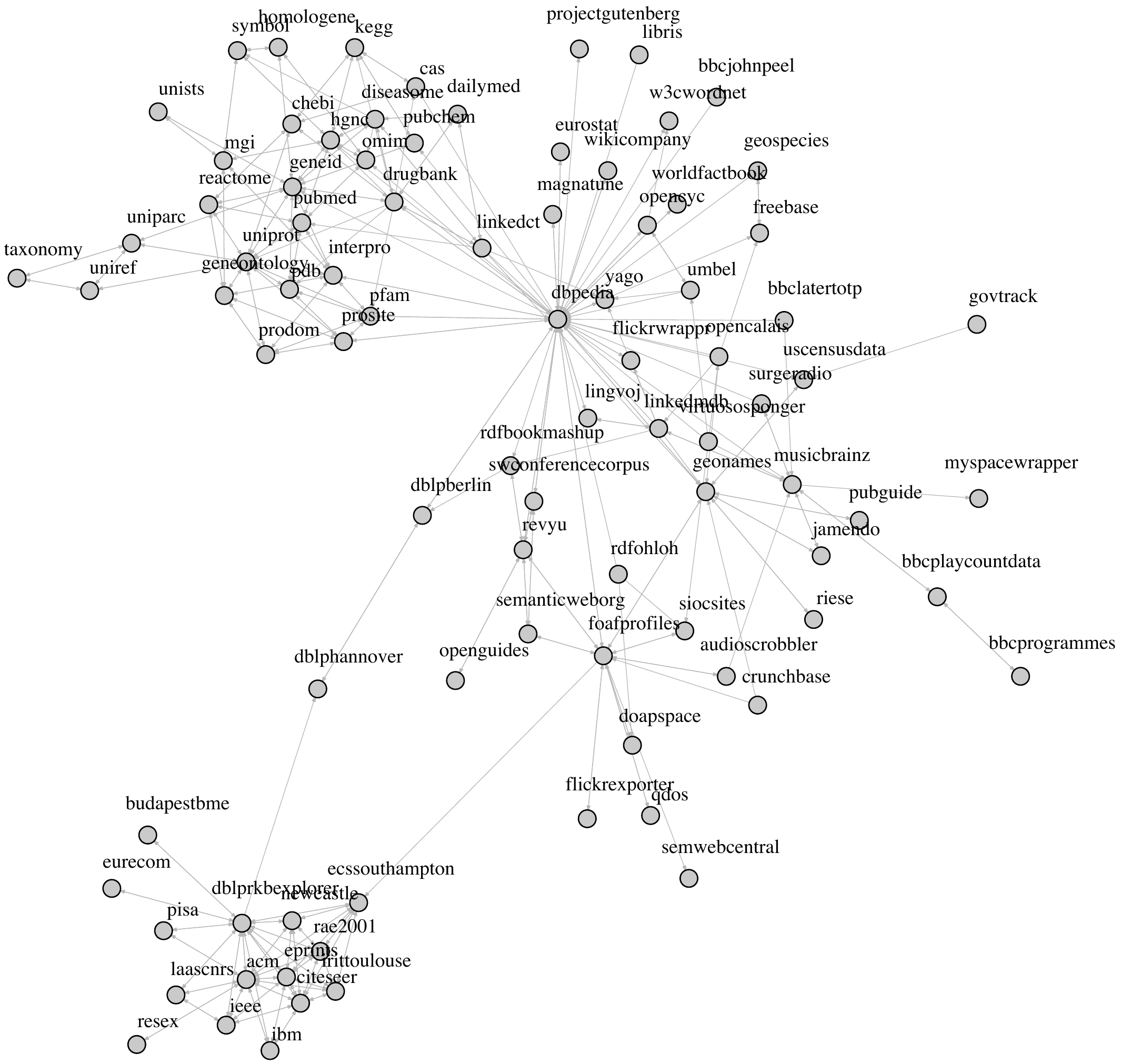}
	\caption{A representation of the $89$ RDF data sets currently in the Linked Data cloud.\label{fig:lod-graph}}
\end{figure*}

Figure \ref{fig:lod-graph} presents a visualization of the linking structure of the $89$ data sets currently in the Linked Data cloud.\footnote{The Linked Data cloud is a subset of the larger Web of Data that includes those data sets that are directly or indirectly connected to DBpedia and are maintained by the Linked Data community.} Each vertex represents a unique data set that exists on an Internet server. The directed relationships denote that the source data set references resources in the sink data set. The current Linked Data cloud maintains approximately $4.5$ billion relationships on data from various domains of interest. Table \ref{tab:lod-categories} indicates the domain of interest for each data set.

\begin{table}[h!]
\caption{\label{tab:lod-categories} The domains of the $89$ data sets currently in the Linked Data cloud.}
\begin{scriptsize}
\begin{center}
\begin{tabular}{ll | ll | ll}
\hline
data set & domain & data set & domain & data set & domain \\
\hline
acm&computer&geospecies &biology &pubchem& biology\\
audioscrobbler&music &govtrack& government &pubguide& books\\
bbcjohnpeel&music &hgnc&biology&pubmed& medical\\
bbclatertotp&music &homologene& biology&qdos& social\\
bbcplaycountdata&music &ibm&computer &rae2001& computer\\
bbcprogrammes&media &ieee&computer &rdfbookmashup& books\\
budapestbme&computer &interpro& biology&rdfohloh& social\\
cas& biology &irittoulouse& computer &reactome&biology\\
chebi&biology &jamendo& music &resex&computer\\
citeseer&computer &kegg& biology&revyu& reference\\
crunchbase&business &laascnrs&computer &riese& government\\
dailymed&medical &libris& books&semanticweborg& computer\\
dblpberlin&computer &lingvoj& reference&semwebcentral& social\\
dblphannover& computer &linkedct& medical &siocsites& social \\
dblprkbexplorer& computer &linkedmdb& movie &surgeradio& music\\
dbpedia&general &magnatune& music &swconferencecorpus& computer\\
diseasome&medical &mgi& biology &symbol& medical \\
doapspace&social &musicbrainz&music &taxonomy& reference \\
drugbank&medical &myspacewrapper&social &umbel& general \\
ecssouthampton&computer &newcastle& computer &uniparc&biology\\
eprints& computer&omim&biology&uniprot& biology\\
eurecom& computer&opencalais& reference&uniref& biology\\
eurostat&government &opencyc& general&unists& biology\\
flickrexporter&images &openguides& reference&uscensusdata&government\\
flickrwrappr&images &pdb& biology&virtuososponger&reference\\
foafprofiles&social &pfam& biology&w3cwordnet&reference\\
freebase&general &pisa& computer&wikicompany&business\\
geneid&biology &prodom& biology&worldfactbook&government\\
geneontology&biology &projectgutenberg& books&yago&general\\
geonames&geographic &prosite&biology&&\\
\hline
\end{tabular}
\end{center}
\end{scriptsize}
\end{table}

By publicly exposing data sets such as Amazon.com's RDF book mashup, MusicBrainz.org's metadata archive, the Internet Movie Database's (IMDB) collection of movie facts, Revyu.com's user ratings, and the publishing and conference behavior of scholars, the Web of Data hosts a rich model of the world that is not built by a single provider, but by many providers collaboratively integrating their data. Such a massive public data structure can be exploited by a community of developers focused on ensuring that the right resource reaches the right person at the right time. Ultimately, an orchestration of this magnitude could yield virtuous individuals whose lives are filled with experiences tailored to their nature.

The Web of Data already includes data sets that are pertinent to modeling individuals and resources; however, the success of a eudaemonic system depends on the availability of data regarding the individual and their past, current, and predicted responses to resources. At the societal level, research has demonstrated that resources relevant to flourishing are those that support life expectancy, nutrition, purchasing power, freedom, equality, education, literacy, access to information, and mental health \cite{progress:heylighen2000}.\footnote{The World Database of Happiness provides data concerning the study of well-being worldwide and is available at  \ttt{http://worlddatabaseofhappiness.eur.nl}.} At the individual level, gathering and maintaining data regarding fluctuations in an individual's well-being in relation to resources would support the automatic determination of optimal future states for that individual.

While the Linked Data community is providing a distributed data structure, they are not providing a distributed process infrastructure \cite{rodriguez:distributed2008}. Currently, the Linked Data practice is to mint \ttt{http}-based URIs. These \ttt{http}-based URIs are dereferenced in order to retrieve a collection of RDF statements associated with that URI. The problem with this model is that it relegates the Web of Data to use primarily by man. For a machine to traverse parts of the larger Web of Data, the pull-based mechanism of HTTP greatly reduces the speed of processing. It would be unfortunate to limit the sophistication of the algorithms that can reasonably process this data due to an infrastructure issue that can be solved using distributed computing.

Ultimately, once these computational hurdles are overcome, what can emerge is a ``society of algorithms" that leverages the Web of Data to support individuals in ways that are not possible given the current recommender system architectures. Through such an undertaking, the niche recommender system is transformed into a eudaemonic system, one that fosters a society of individuals where the vocation one takes, the person one dates, the books one reads, the restaurants one frequents, and so on are chosen not through the advice of one's family, friends, and community, but through a deep computational understanding of what is required for that individual to live an optimal life.

\section{Conclusion}

The evolution of the recommender system to the eudaemonic system will be driven by the public exposure of massive-scale, interlinked, heterogenous data and algorithms that can effectively and efficiently process such data. The goal of a eudaemonic system is to orient people towards those resources that will produce a life that is devoid of pretense, doubt, and ultimately, fear. That is, a eudaemonic system will aid the individual in situating themselves within that area of the world that makes sense to them. A pre-conate eudaemonic system would direct the individual to choose need-mitigating options before the individual becomes aware of their need. In other words, the individual would choose options that they do not perceive as necessary. Without the perception of need, the individual would take on faith that the algorithm knows what is best for them in a resource complex world. Thus, the perfect life is not an aspiration, but a well-computed path.

\section*{Note}

Faith in the Algorithm is a series of articles that focuses on the intersection of political philosophy, ethics, and computation.

\end{document}